\documentclass[aps,preprint,nofootinbib]{revtex4}
\usepackage{amsmath}
\usepackage{graphicx}
\usepackage{color}
\usepackage{slashed}
\usepackage{subfigure}
\usepackage{tabularx}
\usepackage{mathtools}
\usepackage{hyperref}
\usepackage{multirow}
\numberwithin{figure}{section}
\allowdisplaybreaks[4]

\begin{document}

\title{$J/\psi$ Polarization and $p_T$ distribution in $c\ \!\bar{c}$ associated hadroproduction at $\mathcal{O}(\alpha_s^5)$\\[0.7cm]}

\author{\vspace{1cm} Qi-Ming Feng$^{1}$ and Cong-Feng Qiao$^{1,2}$\footnote[1]{qiaocf@ucas.ac.cn}}

\affiliation{\small {$^1$School of Physics, University of Chinese Academy of
Sciences, Beijing 100049, China\\
$^2$ICTP-AP, University of Chinese Academy of Sciences, Beijing 100190, China}}

\author{~\\~\\~\\}

\begin{abstract}
\vspace{0.5cm}
The quarkonium production in association with a heavy-quark pair of the same flavor was found plays an important role in various production schemes, let alone it provides a distinctive signature that could be studied in experiment. Within the NRQCD framework, we perform the first complete calculation of the hadroproduction process $gg\to J/\psi+c\bar{c}$ at the next-to-leading order (NLO) in the expansion of strong coupling constant $\alpha_s$. The result tells that the NLO corrections substantially enhance the color-singlet yield and alter the predicted polarization patterns, and hence lead to a better agreement with experimental data across the $p_T$ spectrum. In contrast to charm-quark fragmentation, the $J/\psi$ and $c\bar{c}$ associated production channel displays a markedly different kinematic behavior, underscoring its distinct role in the production mechanism. While the inclusion of this process greatly reduces the discrepancy between color-singlet theoretical prediction and experimental measurement, a residual gap of approximately one order of magnitude still remains.
\end{abstract}
\maketitle

\newpage

\section{Introduction}

Heavy quarkonium production serves as a unique probe into both perturbative and nonperturbative aspects of QCD, including the mechanisms of hadronization and the nature of color confinement. Despite decades of progress, the quarkonium production mechanism remains an unsettled issue. The nonrelativistic QCD (NRQCD) factorization framework~\cite{Bodwin:1994jh} provides a systematic approach to separate the perturbative caculable short-distance sector from the long-distance hadronization part via a double expansion in the strong coupling constant and heavy-quark velocity. However, studies on the higher order velocity expansion color-octet (CO) contributions show somewhat incompatible results~\cite{Bodwin:2012ft}.

Within NRQCD, the production cross section is expressed as a sum over intermediate quark-antiquark states, each weighted by a short-distance coefficient (SDC) calculable in perturbation theory, and a long-distance matrix element (LDME) that encodes nonperturbative transition probabilities. While SDCs are process dependent, LDMEs are assumed to be universal, allowing global fit across multiple production channels.

Extensive investigations of $J/\psi$ transverse momentum ($p_T$) distributions at colliders have established the essential role of CO mechanisms \cite{Chang:2009uj,Artoisenet:2009xh,Campbell:2007ws,Artoisenet:2007xi,Gong:2008sn,Lansberg:2010vq,Kramer:1994zi,Kramer:1995nb,Butenschoen:2009zy,Butenschoen:2011ks,Butenschoen:2012px,Chao:2012iv,Ma:2010jj,Ma:2010yw,Zhang:2009ym,Gong:2012ug,Brambilla:2024iqg}. However, significant discrepancies persist between theoretical predictions and experimental measurements, especially on charmonium polarization and high-$p_T$ region. The discrepancies can be mainly traced back to the fitting schemes in extracting CO LDMEs. Although the feeddown effects from higher charmonium states, such as $\psi(2S)$ and $\chi_{cJ}$, may help reconcile some of the inconsistencies in the total prompt $J/\psi$ yield \cite{Gong:2012ug}, the determination of a universal and consistent set of CO LDMEs remains an open challenge.

On experimental aspect, the CMS Collaboration reported the measurements of $J/\psi$ polarization at the LHC RUN I with colliding energy $\sqrt{s} = 7\ \mathrm{TeV}$~\cite{CMS:2013gbz} and RUN II with $\sqrt{s} = 13\ \mathrm{TeV}$~\cite{CMS:2024igk}. The observed polar anisotropy parameter $\lambda_\theta$ shows only mild variation with $p_T$ and tends to a value around 0.3 in high-$p_T$ region. This behavior suggests a suppression or cancellation between transverse and longitudinal components, implying the presence of unaccounted production channels with nearly unpolarized characteristics.

In this context, the associated production process $pp \to J/\psi + c\bar{c}$ emerges as a promising candidate. Earlier studies~\cite{Qiao:2003ba,Zhang:2006ay,Artoisenet:2007xi,Gong:2008sn,Chen:2016hju,Feng:2024heh} examined $J/\psi + c\bar{c}$ associated production across various production mechanisms. These works reveal sizable contributions relative to inclusive $J/\psi$ production, particularly in $pp$ collision. Researches also indicate that the NLO corrections to the charmonium production tend to be substantial \cite{Zhang:2006ay,Gong:2008sn,Chen:2016hju,Feng:2024heh}. In particular, the gluon-gluon fusion process was found may yield nearly unpolarized $J/\psi$ mesons \cite{Artoisenet:2007xi,Gong:2008sn}, in agreement with experimental data. Despite its potential importance of the $gg \to J/\psi + c\bar{c}$ process in the color-singlet (CS) channel, nevertheless, the full NLO QCD corrections to it have never been computed, mainly due to the complexity of the multi-body final-state phase space and the large number of contributing Feynman diagrams.

In this Letter, we report the first complete calculation of the $gg \to J/\psi + c\bar{c}$ process at NLO QCD ($\mathcal{O}(\alpha_s^5)$), retaining leading-order accuracy in the NRQCD velocity expansion. We present results for both the $J/\psi$ transverse momentum distribution and polarization, and demonstrate that the NLO corrections significantly enhance the CS contribution and yield polarization behavior in improved agreement with experimental observations.

\section{Calculations}

The spin density matrix $d\sigma_{s_z s_z'}$ characterizes both the production rate and polarization of the produced $J/\psi$. At leading order in the NRQCD velocity expansion, it factorizes as
\begin{align}\label{c_s}
    d\sigma_{s_z s_z'} (pp \to J/\psi + c\bar{c})& = \sum_{i,j} \int  d\eta_i\, d\eta_j\, f_{i/p}(\eta_i,\mu^2) f_{j/p}(\eta_j,\mu^2) \nonumber \\
    & \times d\hat\sigma_{s_z s_z'} (ij \to c\bar c[^3\!S_1] + c\bar{c}) \langle \mathcal{O}^{J/\psi}(^3\!S_1) \rangle\ ,
\end{align}
where $\langle \mathcal{O}^{J/\psi}(^3\!S_1^{[1]}) \rangle$ denotes the CS LDME. The partonic spin density matrix is built from helicity amplitudes via
\begin{align}\label{pcs}
    d\hat{\sigma}_{s_z s'_z} \propto 
    \sum_{s_i, s_j, s_c, s_{\bar{c}}}
    \mathcal{A}^{s_i s_j s_z s_c s_{\bar{c}}} 
    ({\mathcal{A}^{s_i s_j s'_z s_c s_{\bar{c}}}})^* \ .
\end{align}
Here, $\mathcal{A}^{s_i s_j s_z s_c s_{\bar{c}}}$ signifies the helicity amplitude with spin indices $(s_i,s_j,s_z,s_c,s_{\bar c})$ corresponding to the initial partons $i$, $j$, the final $J/\psi$, and the final charm quark and antiquark, respectively. The polarization of the $J/\psi$ is determined by its dilepton decay angular distribution in the $J/\psi$ rest frame, parameterized as~\cite{Braaten:2014ata,Faccioli:2010kd}:
\begin{align}\label{polar_def}
    \frac{d\sigma(J/\psi\to \ell^+\ell^-)}{d\cos\theta} \propto 1 + \lambda_\theta \cos^2 \theta\ ,
\end{align}
where $\lambda_\theta$ characterizes the reconstructed $J/\psi$ polarization, $\lambda_\theta\to \pm 1$ correspond to transverse polarization ($+1$) and longitudinal polarization ($-1$) of $J/\psi$, respectively. It can be expressed in terms of spin density matrix as:
\begin{align}\label{params}
    \lambda_\theta = \frac{d\sigma_{11} - d\sigma_{00}}{d\sigma_{11} + d\sigma_{00}}\ .\quad
\end{align}

\begin{figure}
    \includegraphics[width=\textwidth]{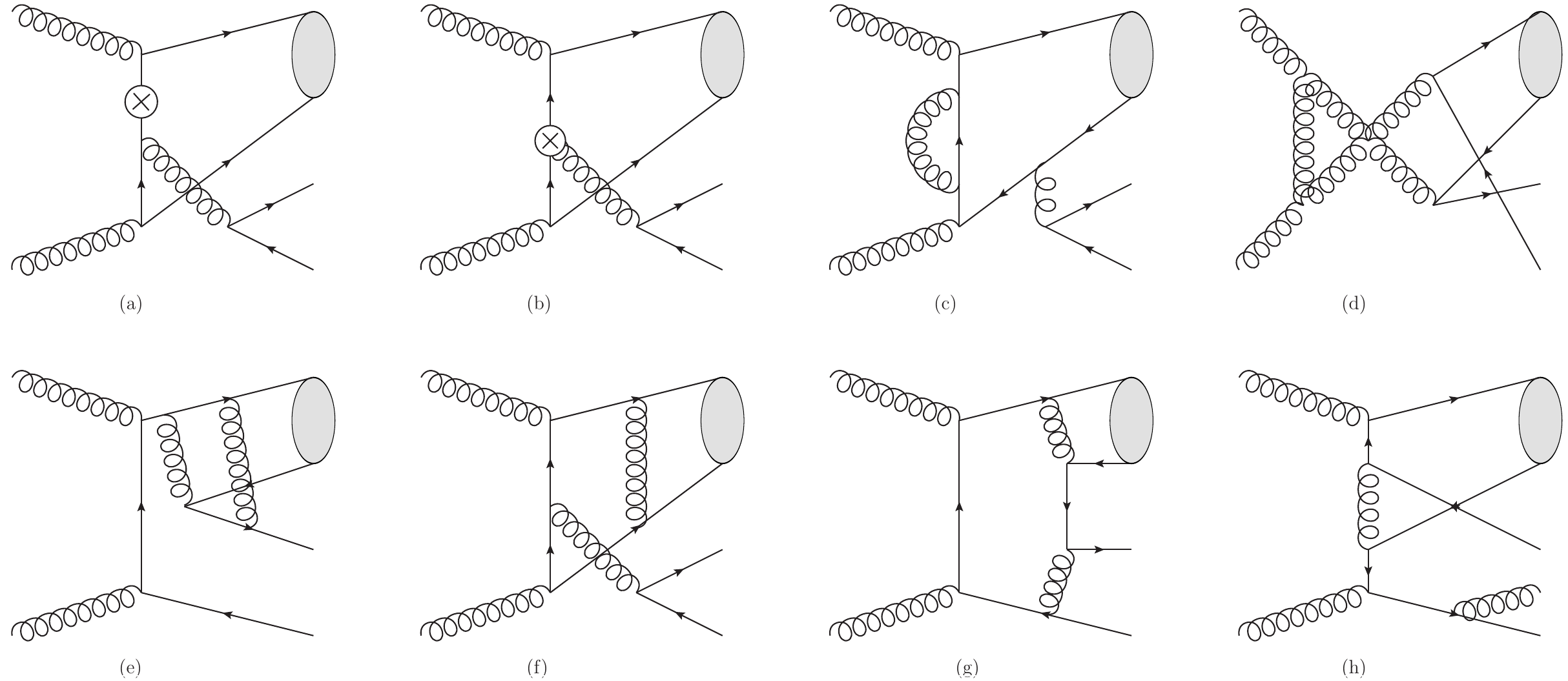}
    \caption{Typical NLO diagrams of $gg\to J/\psi+c\bar{c}$ process. (a) and (b) are counterterm diagrams, (c)$\sim$(g) are 2-6 point one-loop diagrams, incorporating UV, soft, collinear, Coulomb divergences, respectively. (h) represents real correction.}\label{dias}
\end{figure}

The $g\,(k_1)+g\,(k_2)\to J/\psi\,(p_0)+c\,(p_1)+\bar{c}\,(p_2)$ associated process involves 42 LO and over 2,000 NLO Feynman diagrams. Typical NLO Feynman diagrams are shown in Fig.~\ref{dias}. Our calculation largely follows the strategy outlined in~\cite{Feng:2024heh}. Scattering amplitudes are first analytically generated and projected onto the CS $c\bar{c}[^3S_1]$ state. We adopt the spinor helicity formalism to numerically evaluate amplitudes, except for loop diagrams containing Coulomb divergences. For the latter, we apply the conventional squared-amplitude method to compute both the total and the longitudinal $J/\psi$ cross sections. The longitudinal polarization cross section $d\sigma_L = d\sigma_{00}$ is evaluated by employing the longitudinal polarization projection tensor
\begin{align}\label{longivec}
\Pi_L^{\mu\nu} = \varepsilon_L^\mu(p_0)\varepsilon_L^{*,\nu}(p_0) = \frac{1}{M^2} \left(p_0^\mu - \frac{M^2}{k_1 \cdot p_0} k_1^\mu\right) \left(p_0^\nu - \frac{M^2}{k_1 \cdot p_0} k_1^\nu\right)\ ,
\end{align}
where $M$ is the mass of the $J/\psi$. The transverse cross section is obtained via $d\sigma_{T}=d\sigma_{11}+d\sigma_{-1,-1}=d\sigma-d\sigma_{L}$, where $d\sigma$ is the total cross section.

Loop integrals without Coulomb singularities are evaluated by using integrand reduction via the Laurent-expansion method~\cite{Mastrolia:2012bu,Peraro:2014cba}, while those terms containing Coulomb poles are handled via integration-by-parts (IBP) reduction using \texttt{NeatIBP}~\cite{Wu:2023upw}. For loop diagrams, the calculation may be greatly simplified through subtracting those contributions related by symmetric permutations of the initial gluons and the final (anti-)charm quarks. Ultraviolet (UV) divergences are canceled via renormalization. The renormalization constant $Z_g$ is defined in the modified minimal subtraction ($\overline{\rm MS}$) scheme, while $Z_2$, $Z_m$, and $Z_3$ are defined in the on-shell (OS) scheme. Infrared (IR) divergences arising in the real corrections are extracted by means of the two-cutoff phase space slicing method \cite{Harris:2001sx}. Thus, the total cross section at NLO accuracy can be expressed as:
\begin{align}\label{totcs}
    \sigma_{\rm{tot}} = \int_{\rm{3-body}} \left( d\sigma^{\rm{LO}} + d\sigma^{\rm{Virtual}} + d\sigma^{\rm{Real}}_{\rm{S}} + d\sigma^{\rm{Real}}_{\rm{HC}} \right) + \int_{\rm{4-body}} d\sigma^{\rm{Real}}_{\rm{H\overline{C}}} \ .
\end{align}
Here, $d\sigma^{\rm{LO}}$, $d\sigma^{\rm{Virtual}}$, and $d\sigma^{\rm{Real}}$ denote the LO, virtual, and real contributions, respectively. The virtual term comprises three separately computed contributions: counterterms from renormalization, loop diagrams with Coulomb singularities, and those without. The real contributions are partitioned into soft ($\rm S$), hard collinear ($\rm HC$), and hard non-collinear ($\rm H\overline{C}$) regions.

As an analytic check, the IR divergences of the loop amplitudes are extracted following the method of~\cite{Dittmaier:2003bc}. The cancellation between virtual and real corrections leads to
\begin{align}
    2\, \mathcal{A}_{\rm div}^{\rm Virtual}(\mathcal{A}^{\rm LO})^* = -|\mathcal{A}_{\rm div}^{\rm Real}|^2 = \sum_{i,j=1}^{3} \hat{\mathcal{A}}_i^{\rm LO} \, \boldsymbol{\Delta}^{ij} \, (\hat{\mathcal{A}}_j^{\rm LO})^* \ .
\end{align}
Here, $\hat{\mathcal{A}}_i^{\rm LO}$ represent the color-decomposed Born-level amplitudes:
\begin{align}
    \mathcal{A}^{\rm LO} = \delta^{g_1 g_2} \delta_{c\bar{c}}\, \hat{\mathcal{A}}_{1}^{\rm LO} 
    + (\mathcal{T}^{g_1} \mathcal{T}^{g_2})_{c\bar{c}}\, \hat{\mathcal{A}}_{2}^{\rm LO} 
    + (\mathcal{T}^{g_2} \mathcal{T}^{g_1})_{c\bar{c}}\, \hat{\mathcal{A}}_{3}^{\rm LO}
\end{align}
with $g_{1,2}$ and $c\bar{c}$ denoting the color indices of the initial gluons and final-state charm quarks, respectively. The IR-divergent kernel $\boldsymbol{\Delta}$ reads
\begin{align}
    \boldsymbol{\Delta} = \resizebox{0.85\textwidth}{!}{
$\displaystyle
\left(\begin{array}{ccc}
-\frac{144}{\varepsilon^2} + \frac{8(4\mathcal{C}_0-35)}{\varepsilon} & -\frac{24}{\varepsilon^2} + \frac{2(8\mathcal{C}_0+9(\mathcal{C}_1-\mathcal{C}_2-\mathcal{C}_3+\mathcal{C}_4)-70)}{3\varepsilon} & -\frac{24}{\varepsilon^2} + \frac{2(8\mathcal{C}_0-9(\mathcal{C}_1-\mathcal{C}_2-\mathcal{C}_3+\mathcal{C}_4)-70)}{3\varepsilon} \\[2pt]
-\frac{24}{\varepsilon^2} + \frac{2(8\mathcal{C}_0+9(\mathcal{C}_1-\mathcal{C}_2-\mathcal{C}_3+\mathcal{C}_4)-70)}{3\varepsilon} & -\frac{32}{\varepsilon^2} + \frac{\mathcal{C}_0+72\mathcal{C}_1-9\mathcal{C}_2-9\mathcal{C}_3+72\mathcal{C}_4-560}{9\varepsilon} & \frac{4}{\varepsilon^2} + \frac{10\mathcal{C}_0-9(\mathcal{C}_1+\mathcal{C}_2+\mathcal{C}_3+\mathcal{C}_4)+70}{9\varepsilon} \\[2pt]
-\frac{24}{\varepsilon^2} + \frac{2(8\mathcal{C}_0-9(\mathcal{C}_1-\mathcal{C}_2-\mathcal{C}_3+\mathcal{C}_4)-70)}{3\varepsilon} & \frac{4}{\varepsilon^2} + \frac{10\mathcal{C}_0-9(\mathcal{C}_1+\mathcal{C}_2+\mathcal{C}_3+\mathcal{C}_4)+70}{9\varepsilon} & -\frac{32}{\varepsilon^2} + \frac{\mathcal{C}_0-9\mathcal{C}_1+72\mathcal{C}_2+72\mathcal{C}_3-9\mathcal{C}_4-560}{9\varepsilon}
\end{array}\right)$
    }.
\end{align}
Here, 
\begin{align}
    &\mathcal{C}_0 = \frac{4(s_{45}-2m_c^2)}{\sqrt{s_{45}(s_{45}-4m_c^2)}} \ln\left( \frac{\sqrt{s_{45}}+\sqrt{s_{45}-4m_c^2}}{2m_c} \right),\quad 
    \mathcal{C}_1 = \ln\left( \frac{(m_c^2 - t_{14})^2}{m_c^2 s_{12}} \right), \nonumber\\
    &\mathcal{C}_2 = \ln\left( \frac{(m_c^2 - t_{15})^2}{m_c^2 s_{12}} \right),\quad 
    \mathcal{C}_3 = \ln\left( \frac{(m_c^2 - t_{24})^2}{m_c^2 s_{12}} \right),\quad 
    \mathcal{C}_4 = \ln\left( \frac{(m_c^2 - t_{25})^2}{m_c^2 s_{12}} \right)
\end{align}
with $s_{ij}$ and $t_{ij}$ being the corresponding Mandelstam invariants.

\section{Numerical results}

In numerical calculation, the charm quark mass is taken as half of the $J/\psi$ mass, i.e. $m_c = 1.5\ \mathrm{GeV}$. The renormalization and factorizaiton scales are set as $\mu_r=\mu_f=m_T\equiv\sqrt{p_T^2+4m_c^2}$, where $m_T$ is the transverse mass of the $J/\psi$. The running coupling constant is evaluated at one-loop (two-loop) accuracy for LO (NLO) calculation, and the proton parton distribution functions (PDFs) CTEQ6L1 (CTEQ6M)~\cite{Pumplin:2002vw} are employed at LO (NLO). The CS LDME satisfies $\langle \mathcal{O}^{J/\psi}(^3\!S_1) \rangle = \frac{9}{2\pi}|R(0)|^2$, where the squared radial wave function at the origin, $|R(0)|^2$, is extracted form the leptonic decay width $\Gamma(J/\psi\to e^+e^-)=5.55\ \rm{keV}$~\cite{ParticleDataGroup:2020ssz}. 

To isolate direct $J/\psi$ production from prompt experimental data, feeddown contributions from $\psi(2S)$ and $\chi_{cJ}$ states are included. The $\psi(2S)$ feeddown is fitted up to $120\ \rm{GeV}$ using ATLAS data~\cite{ATLAS:2023qnh}, with branching ratios $\mathcal{B}(J/\psi \to \mu^+ \mu^-) = 0.593$, $\mathcal{B}(\psi(2S) \to J/\psi) = 0.595$, and $\mathcal{B}(\psi(2S) \to \mu^+ \mu^-) = 0.077$. The transverse momentum of the feeddown $J/\psi$ is approximated by $p_T^{J/\psi} = \frac{m_{J/\psi}}{m_{\psi(2S)}} p_T^{\psi(2S)}$. The $\chi_{cJ}$ feeddown contribution, taken from~\cite{ATLAS:2014ala}, amounts to approximately 26\%. 

\begin{figure}
    \centering
    \subfigure[]{
    \includegraphics[width=0.32\textwidth]{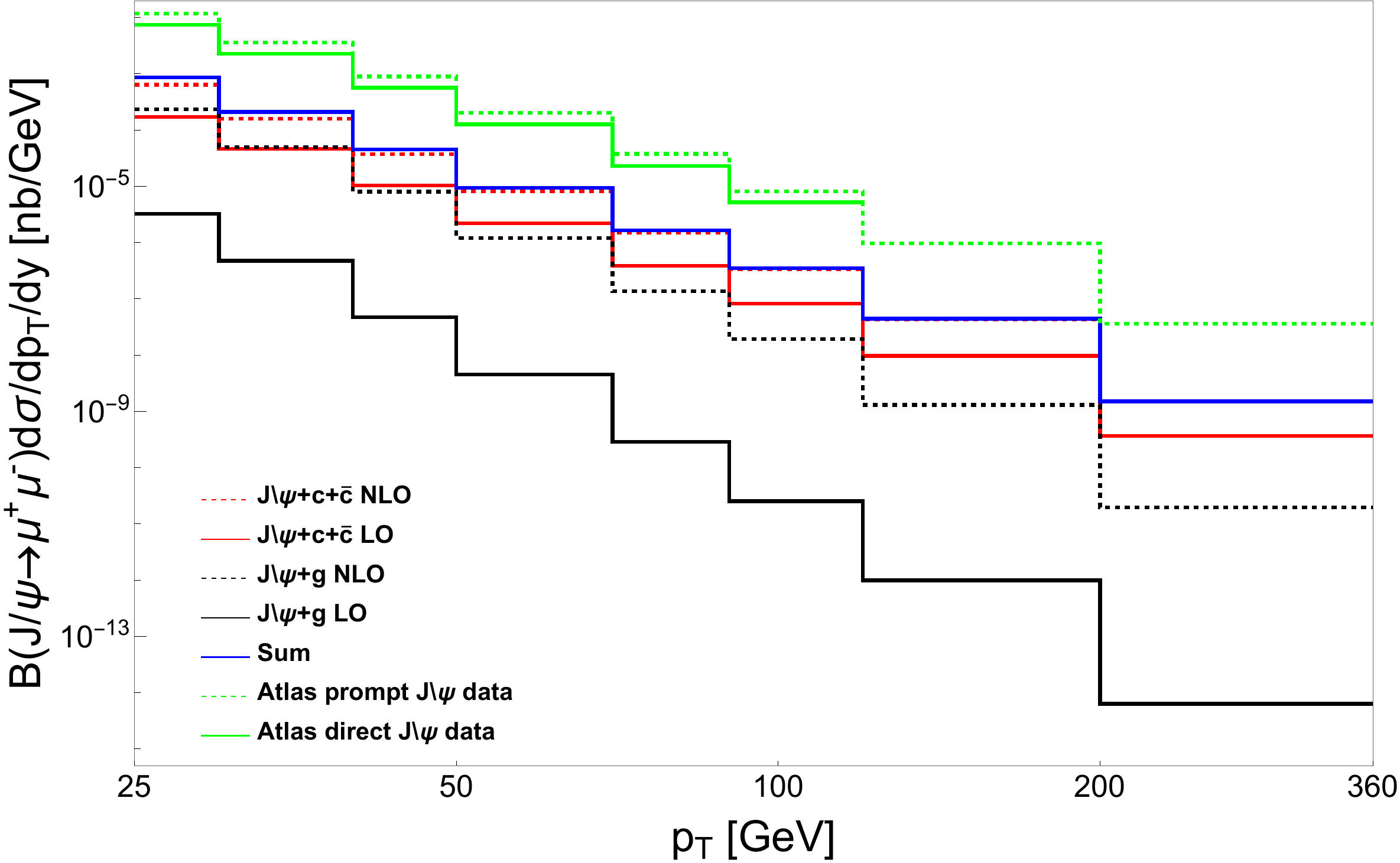}}
    \subfigure[]{
    \includegraphics[width=0.32\textwidth]{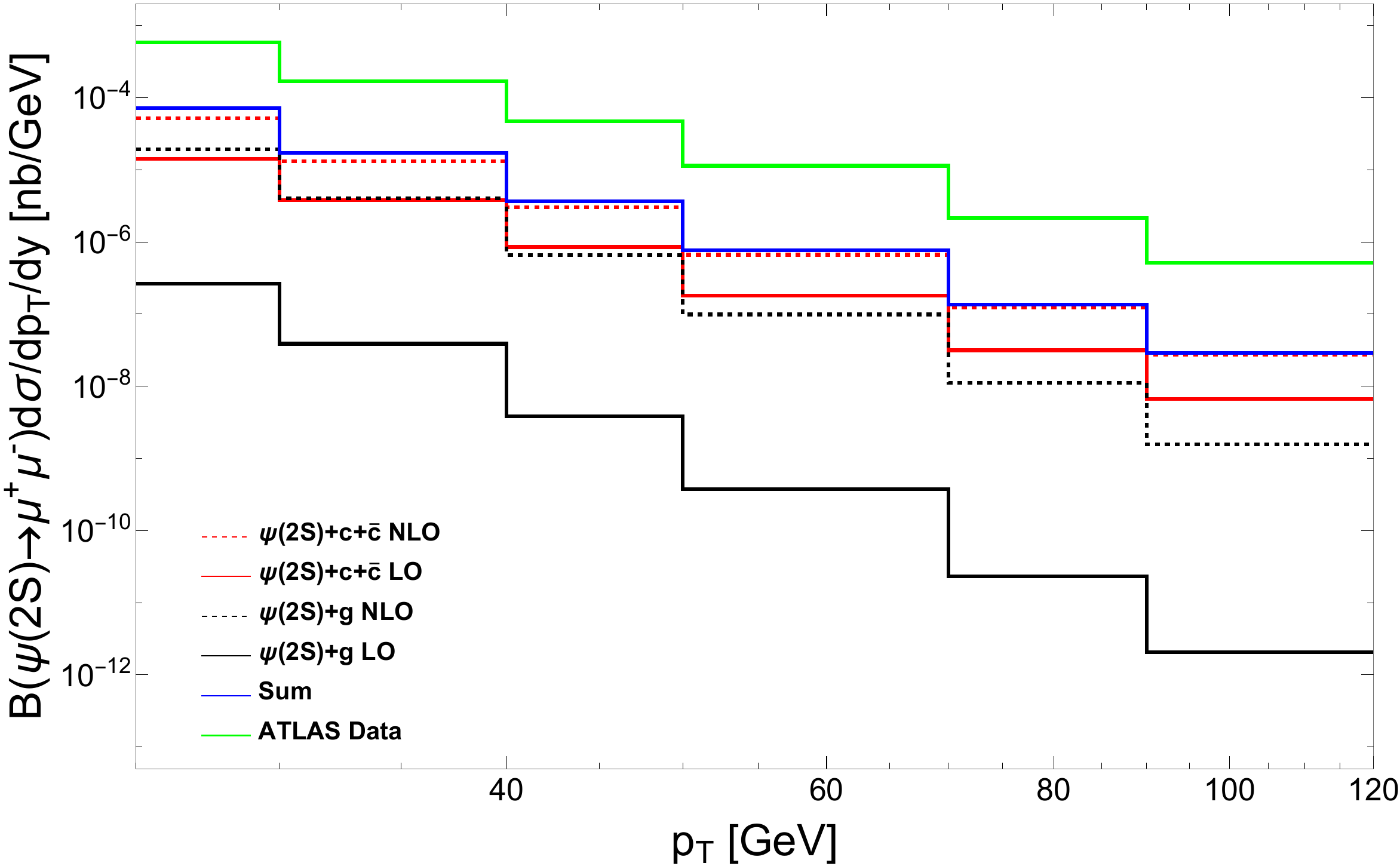}}
    \subfigure[]{
    \includegraphics[width=0.32\textwidth]{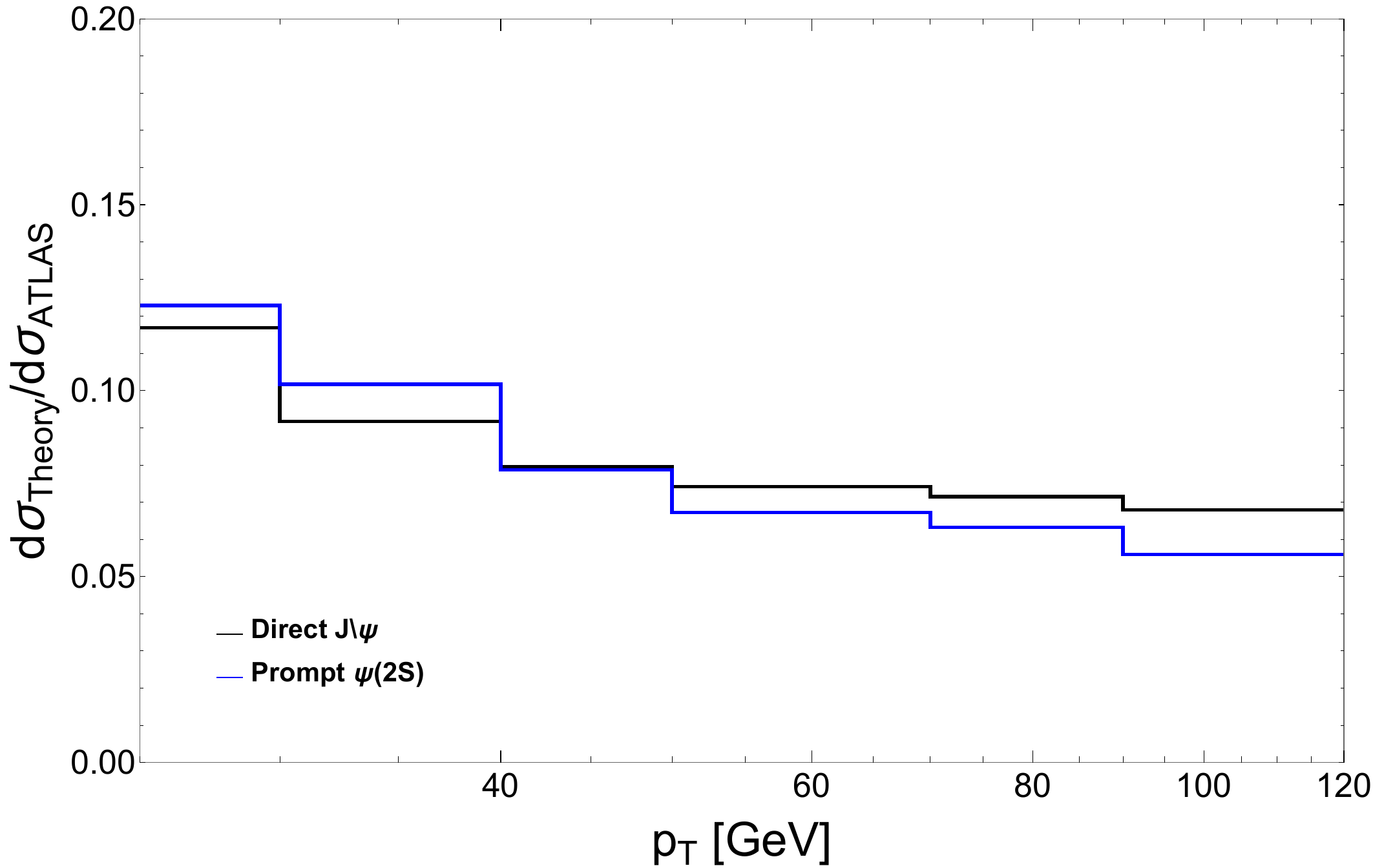}}
    \caption{At $\sqrt{s}=13\ \rm{TeV}$ LHC, differential cross section distributions in $p_{T}$ for (a) $J/\psi$ and (b) $\psi(2S)$ production at LO and NLO in the CS contribution.\@ (c) The ratios of the direct $J/\psi$ and prompt $\psi(2S)$ cross sections over the ATLAS data.}\label{dists}
\end{figure}

FIG.~\ref{dists} presents the $p_T$ differential cross sections at $\sqrt{s} = 13\ \mathrm{TeV}$ LHC experiment. Panel (a) shows the $J/\psi$ production results. For comparison, we also show the LO and NLO CS contributions of the $J/\psi + g$ production process, which dominates inclusive $J/\psi + X$ production. To facilitate comparison, ATLAS prompt $J/\psi$ data~\cite{ATLAS:2023qnh} are rebinned to match our binning, and direct $J/\psi$ production is obtained by subtracting feeddown contributions. Our results indicate that, despite being higher order in $\alpha_s$, the CS contribution from associated $J/\psi + c\bar{c}$ production dominates over $J/\psi + g$, exhibiting a milder falloff with the increase of $p_T$. The shape of the CS $J/\psi + c\bar{c}$ process matches the direct $J/\psi$ data well, though a normalization gap remains.

Given the negligible feeddown to $\psi(2S)$, its production provides a clean probe of short-distance dynamics. FIG.~\ref{dists}(b) shows the $p_T$ distribution of $\psi(2S)$, where the difference from $J/\psi$ in the CSM arises solely from the LDME; we use $|R_{\psi(2S)}|^2 = 0.639\ \mathrm{GeV}^3$ from~\cite{Zhang:2008gp}. Panels (c) displays the ratios of the direct $J/\psi$ and prompt $\psi(2S)$ cross sections to the ATLAS data. The similar $p_T$ dependence of these ratios suggests analogous behavior in the CS channel.

\begin{figure}
    \includegraphics[width=0.7\textwidth]{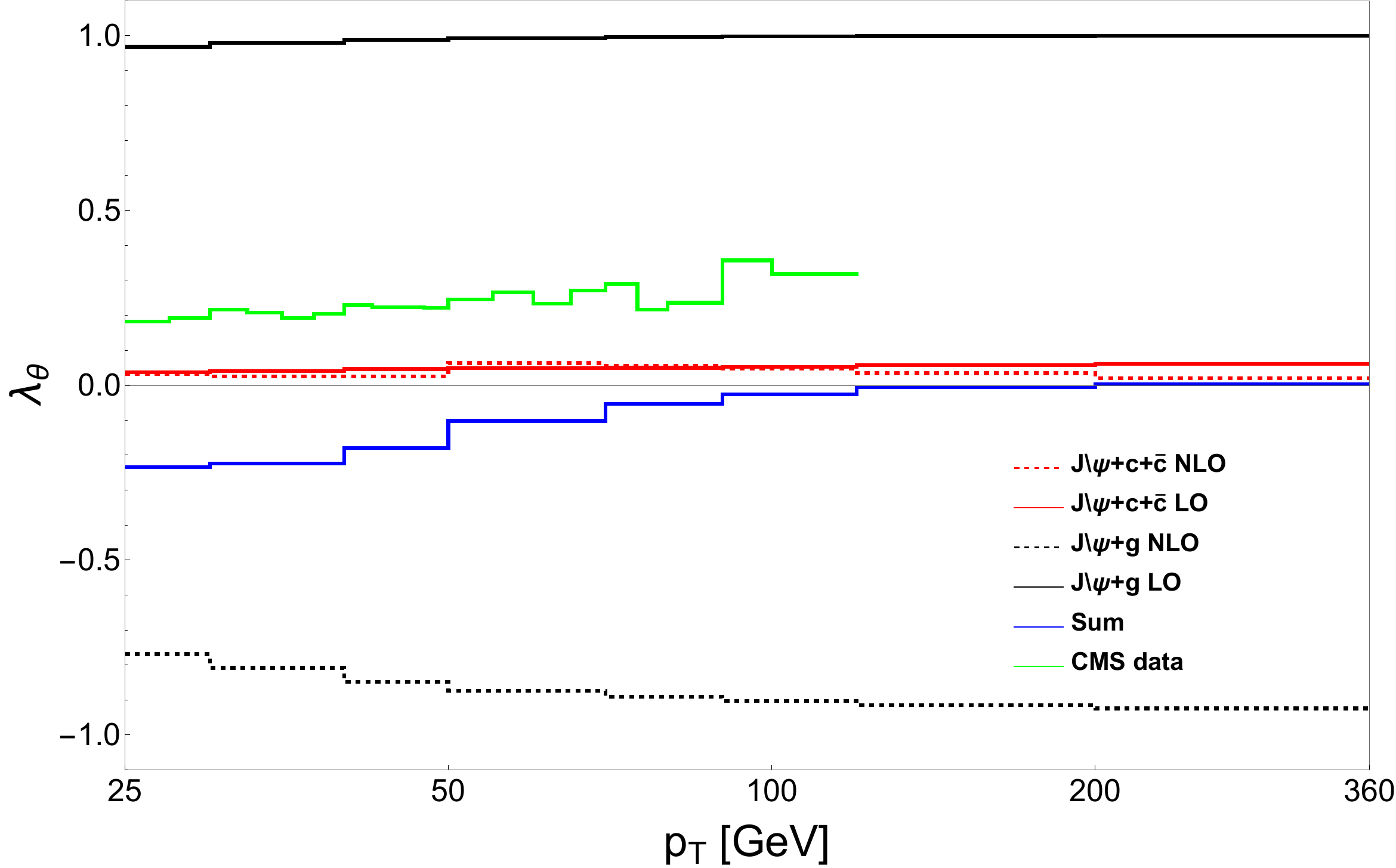}
    \caption{$J/\psi$ polarization parameter $\lambda_\theta$ as a function of $p_T$ in the HX frame at $\sqrt{s} = 13\ \mathrm{TeV}$ LHC.}\label{polar}
\end{figure}

The $J/\psi$ polarization parameters are shown in FIG.~\ref{polar}. Unlike the longitudinal polarization observed in $J/\psi + g$ production, the $J/\psi + c\bar{c}$ channel yields nearly unpolarized $J/\psi$ at both LO and NLO, improving agreement with experimental measurements. The polarization tends toward unpolarized as $p_T$ increases.

\begin{figure}[htb]
    \centering
    \subfigure[]{
    \includegraphics[width=0.45\textwidth]{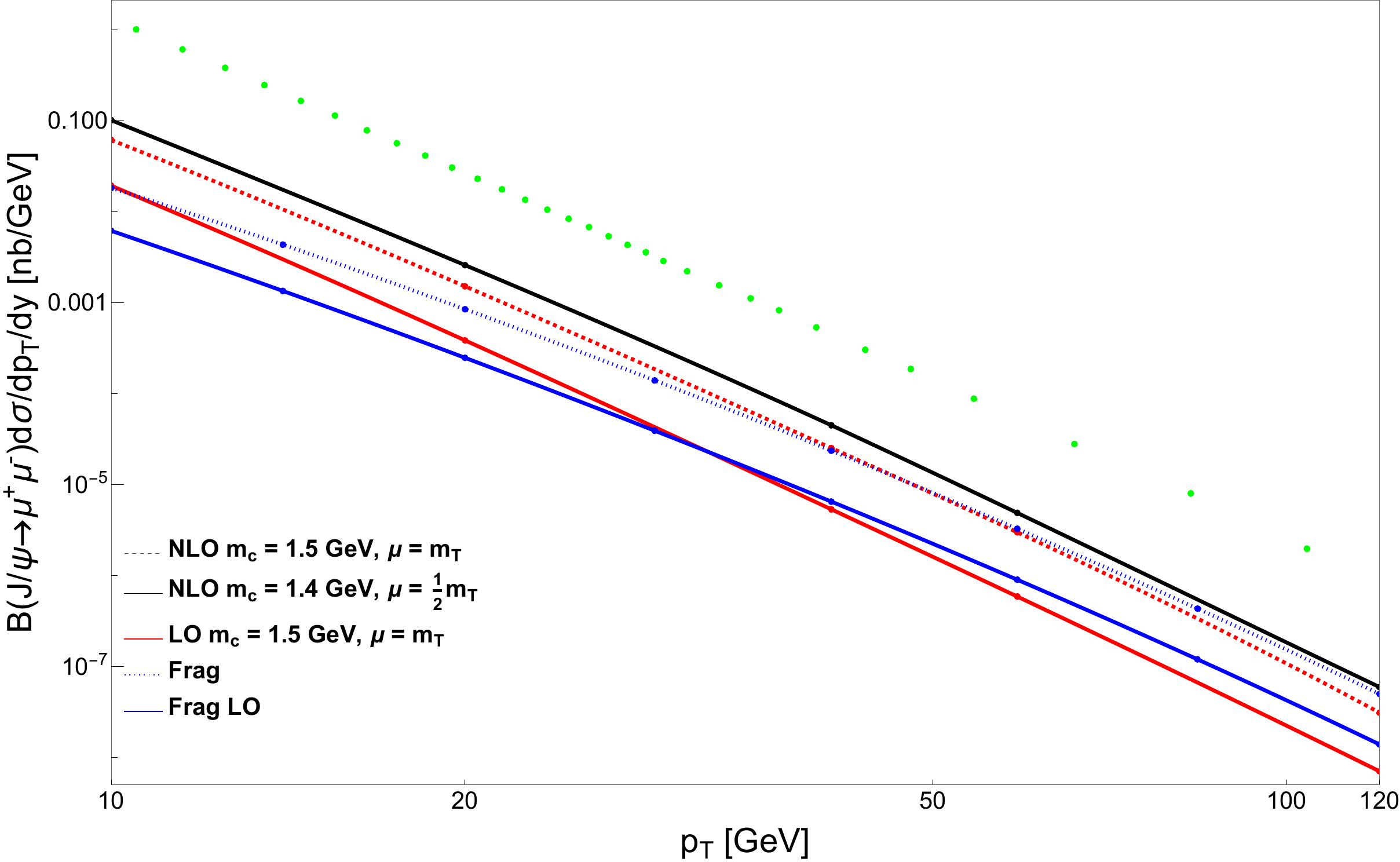}}
    \subfigure[]{
    \includegraphics[width=0.45\textwidth]{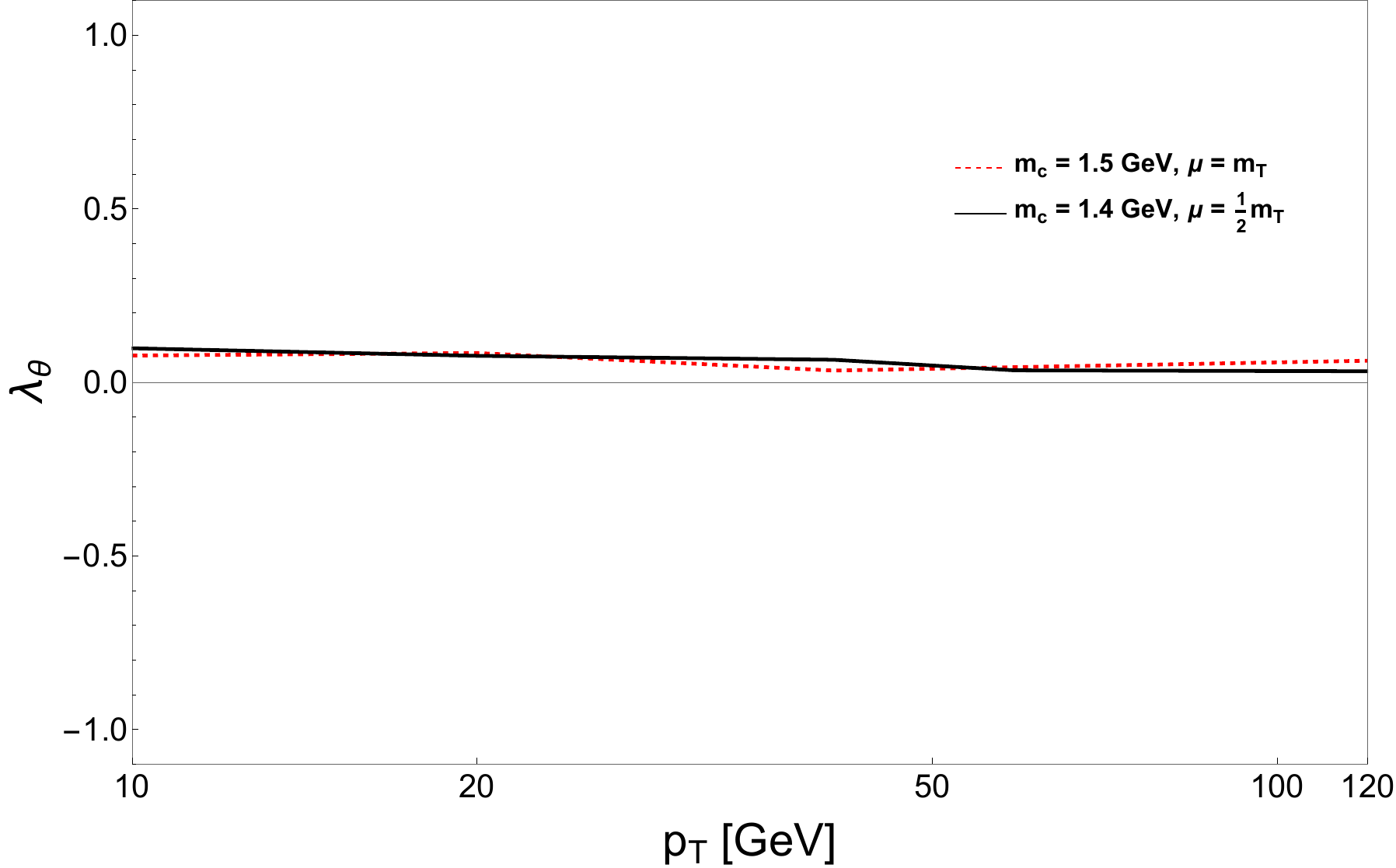}}
    \caption{At $\sqrt{s} = 7\ \mathrm{TeV}$ LHC, (a) $p_T$ differential cross sections for $J/\psi$ production. The green dots are CMS data. (b) $J/\psi$ polarization in $J/\psi + c\bar{c}$ production with varied charm mass and scale.}\label{dist7tev}
\end{figure}

The fragmentation process is expected to contribute comparably to $J/\psi + c\bar{c}$ production. Since charm-quark fragmentation dominates among CS fragmentation contributions, we estimate the fragmentation contribution from the dominant $gg\to c\bar{c}$ process in NLO accuracy:
\begin{align}
    d\sigma_{\rm{frag}}^{\rm{NLO}}(p_0) = 2\left(d\hat\sigma_{gg\to c\bar{c}}^{\rm{NLO}}(\frac{p_0}{z},\mu_f)\otimes D^{\rm{LO}}_{c\to J/\psi}(z,\mu_f) + d\hat\sigma_{gg\to c\bar{c}}^{\rm{LO}}(\frac{p_0}{z},\mu_f)\otimes D^{\rm{NLO}}_{c\to J/\psi}(z,\mu_f)\right)\ .
\end{align} 
Here, $\otimes$ denotes convolution production over $z$. $D^{\rm{LO},\rm{NLO}}_{c\to J/\psi}(z,\mu_f)$ represents the charm to $J/\psi$ fragmentation function at LO and NLO, respectively. The overall factor of 2 accounts for identical $c$ and $\bar{c}$ contributions. The fragmentation functions are taken from~\cite{Braaten:1993mp,Zheng:2019dfk}, and evolved via DGLAP evquation from the initial scale $\mu_0 = 3\, m_c$:
\begin{align}
    \mu_f^2\frac{d}{d \mu_f^2}D^{\rm{LO},\rm{NLO}}_{c\to J/\psi}(z,\mu_f)=\frac{\alpha_s(\mu_f)}{2\pi}P_{cc}(z)\otimes D^{\rm{LO},\rm{NLO}}_{c\to J/\psi}(z,\mu_f)\ ,
\end{align}
where $P_{cc}(z)$ is the regularized splitting function. FIG.~\ref{dist7tev}(a) compares the $J/\psi + c\bar{c}$ hadroproduction, charm-quark fragmentation, and CMS data at $\sqrt{s} = 7\ \mathrm{TeV}$ LHC. At low $p_T$, the associated production dominates over fragmentation at both LO and NLO, in agreement with Ref.\cite{Artoisenet:2007xi}. At high $p_T$, the fragmentation contribution becomes larger than the associated one for $m_c=1.5$ GeV and $\mu=m_T$, consistent with~\cite{Gong:2011et} depicted. As $p_T$ increases, the NLO fragmentation contribution approaches the result with $m_c=1.4\ \mathrm{GeV}$, $\mu = \frac{1}{2} m_T$, exhibiting a clear upward trend that suggests it may overtake the associated process at even higher $p_T$. It indicates that the fragmentation contribution is not a well description of the associated process, instead, their $p_T$ distributions have totally different behaviours. Considering mass and scale uncertainties, the cross section increases roughly twofold, yet a gap with experimental data persists. FIG.~\ref{dist7tev}(b) shows the polarization parameter with varied charm mass and scale, indicating little sensitivity. 

\section{Conclusions}

The quarkonium production in association with a heavy-quark pair of the same flavor turn out to be the dominant processes among various color-singlet production schemes, which bear evident signatures could be studied in experiment.  We present in this work the first calculation of $J/\psi$ hadroproduction associated with a $c\bar{c}$ pair at the NLO QCD in the framework of NRQCD. Results show that the NLO QCD corrections are substantial, may enhance the LO contribution by several times. This confirms the persistence of sizable color-singlet contributions beyond the LO.

According to the calculation, the predicted $p_T$ spectrum of the associated production process exhibits a similar shape to the experimental data, though an obvious discrepancy in magnitude remains. The polarization of $J/\psi$ in the associated production behaves approximately unpolarized in huge transverse momentum scope, in consistent with the recent CMS measurement. The inclusion of the NLO corrections of the concerned process may greatly reduce the uncertainty in the global fitting of color-octet LDMEs. 

Last, it is worth noting that we have compared our calculation with the charm quark fragmentation approximation, and find that the full results are distinct from the latter in both $p_T$ behavior and overall magnitude.

%%%%%%%%%%%%%%%%%%%%%%%%%%%%%%%%%%%%%%%%%%%%%%%%%%%%%%%%%%%%%%%%%%%%%
\vspace{0.3cm} {\bf Acknowledgments}
We are grateful to the High Performance Computing Center (HPCC) of ICTP-AP and Yu-Yang Zhang for the support of computing power.
This work was supported in part by National Natural Science Foundation of China(NSFC) under the Grants 12475087 and 12235008.
%%%%%%%%%%%%%%%%%%%%%%%%%%%%%%%%%%%%%%%%%%%%%%%%%%%%%%%%%%%%%%%%%%%%


\begin{thebibliography}{9}

%\cite{Bodwin:1994jh}
\bibitem{Bodwin:1994jh}
G.~T.~Bodwin, E.~Braaten and G.~P.~Lepage,
%``Rigorous QCD analysis of inclusive annihilation and production of heavy quarkonium,''
Phys. Rev. D \textbf{51} (1995), 1125-1171
[erratum: Phys. Rev. D \textbf{55} (1997), 5853]
%doi:10.1103/PhysRevD.55.5853
[arXiv:hep-ph/9407339 [hep-ph]].
%2648 citations counted in INSPIRE as of 09 Apr 2022

%\cite{Bodwin:2012ft}
\bibitem{Bodwin:2012ft}
G.~T.~Bodwin,
%``Theory of Charmonium Production,''
[arXiv:1208.5506 [hep-ph]].
%15 citations counted in INSPIRE as of 20 Feb 2025

%\cite{Chang:2009uj}
\bibitem{Chang:2009uj}
C.~H.~Chang, R.~Li and J.~X.~Wang,
%``J/psi polarization in photo-production up-to the next-to-leading order of QCD,''
Phys. Rev. D \textbf{80}, 034020 (2009)
%doi:10.1103/PhysRevD.80.034020
[arXiv:0901.4749 [hep-ph]].
%48 citations counted in INSPIRE as of 09 Mar 2024

%\cite{Artoisenet:2009xh}
\bibitem{Artoisenet:2009xh}
P.~Artoisenet, J.~M.~Campbell, F.~Maltoni and F.~Tramontano,
%``J/psi production at HERA,''
Phys. Rev. Lett. \textbf{102}, 142001 (2009)
%doi:10.1103/PhysRevLett.102.142001
[arXiv:0901.4352 [hep-ph]].
%68 citations counted in INSPIRE as of 09 Mar 2024

%\cite{Campbell:2007ws}
\bibitem{Campbell:2007ws}
J.~M.~Campbell, F.~Maltoni and F.~Tramontano,
%``QCD corrections to J/psi and Upsilon production at hadron colliders,''
Phys. Rev. Lett. \textbf{98}, 252002 (2007)
%doi:10.1103/PhysRevLett.98.252002
[arXiv:hep-ph/0703113 [hep-ph]].
%261 citations counted in INSPIRE as of 09 Mar 2024

%\cite{Lansberg:2010vq}
\bibitem{Lansberg:2010vq}
J.~P.~Lansberg,
%``QCD corrections to J/psi polarisation in pp collisions at RHIC,''
Phys. Lett. B \textbf{695}, 149-156 (2011)
%doi:10.1016/j.physletb.2010.10.054
[arXiv:1003.4319 [hep-ph]].
%68 citations counted in INSPIRE as of 09 Mar 2024

%\cite{Artoisenet:2007xi}
\bibitem{Artoisenet:2007xi}
P.~Artoisenet, J.~P.~Lansberg and F.~Maltoni,
%``Hadroproduction of $J/\psi$ and $\Upsilon$ in association with a heavy-quark pair,''
Phys. Lett. B \textbf{653}, 60-66 (2007)
%doi:10.1016/j.physletb.2007.04.031
[arXiv:hep-ph/0703129 [hep-ph]].
%159 citations counted in INSPIRE as of 19 May 2025

%\cite{Gong:2008sn}
\bibitem{Gong:2008sn}
B.~Gong and J.~X.~Wang,
%``Next-to-leading-order QCD corrections to $J/\psi$ polarization at Tevatron and Large-Hadron-Collider energies,''
Phys. Rev. Lett. \textbf{100}, 232001 (2008)
%doi:10.1103/PhysRevLett.100.232001
[arXiv:0802.3727 [hep-ph]].
%174 citations counted in INSPIRE as of 09 Mar 2024

%\cite{Kramer:1994zi}
\bibitem{Kramer:1994zi}
M.~Kramer, J.~Zunft, J.~Steegborn and P.~M.~Zerwas,
%``Inelastic J / psi photoproduction,''
Phys. Lett. B \textbf{348}, 657-664 (1995)
%doi:10.1016/0370-2693(95)00155-E
[arXiv:hep-ph/9411372 [hep-ph]].
%131 citations counted in INSPIRE as of 09 Mar 2024

%\cite{Kramer:1995nb}
\bibitem{Kramer:1995nb}
M.~Kr\"amer,
%``QCD corrections to inelastic J / psi photoproduction,''
Nucl. Phys. B \textbf{459}, 3-50 (1996)
%doi:10.1016/0550-3213(95)00568-4
[arXiv:hep-ph/9508409 [hep-ph]].
%238 citations counted in INSPIRE as of 09 Mar 2024

%\cite{Butenschoen:2009zy}
\bibitem{Butenschoen:2009zy}
M.~Butenschoen and B.~A.~Kniehl,
%``Complete next-to-leading-order corrections to J/psi photoproduction in nonrelativistic quantum chromodynamics,''
Phys. Rev. Lett. \textbf{104}, 072001 (2010)
%doi:10.1103/PhysRevLett.104.072001
[arXiv:0909.2798 [hep-ph]].
%104 citations counted in INSPIRE as of 15 Mar 2024

%\cite{Butenschoen:2012px}
\bibitem{Butenschoen:2012px}
M.~Butenschoen and B.~A.~Kniehl,
%``J/psi polarization at Tevatron and LHC: Nonrelativistic-QCD factorization at the crossroads,''
Phys. Rev. Lett. \textbf{108}, 172002 (2012)
%doi:10.1103/PhysRevLett.108.172002
[arXiv:1201.1872 [hep-ph]].
%235 citations counted in INSPIRE as of 09 Mar 2024

%\cite{Butenschoen:2011ks}
\bibitem{Butenschoen:2011ks}
M.~Butenschoen and B.~A.~Kniehl,
%``Probing nonrelativistic QCD factorization in polarized $J/\psi$ photoproduction at next-to-leading order,''
Phys. Rev. Lett. \textbf{107}, 232001 (2011)
%doi:10.1103/PhysRevLett.107.232001
[arXiv:1109.1476 [hep-ph]].
%58 citations counted in INSPIRE as of 09 Mar 2024

%\cite{Chao:2012iv}
\bibitem{Chao:2012iv}
K.~T.~Chao, Y.~Q.~Ma, H.~S.~Shao, K.~Wang and Y.~J.~Zhang,
%``$J/\psi$ Polarization at Hadron Colliders in Nonrelativistic QCD,''
Phys. Rev. Lett. \textbf{108}, 242004 (2012)
%doi:10.1103/PhysRevLett.108.242004
[arXiv:1201.2675 [hep-ph]].
%289 citations counted in INSPIRE as of 09 Mar 2024

%\cite{Ma:2010jj}
\bibitem{Ma:2010jj}
Y.~Q.~Ma, K.~Wang and K.~T.~Chao,
%``A complete NLO calculation of the $J/\psi$ and $\psi'$ production at hadron colliders,''
Phys. Rev. D \textbf{84}, 114001 (2011)
%doi:10.1103/PhysRevD.84.114001
[arXiv:1012.1030 [hep-ph]].
%123 citations counted in INSPIRE as of 09 Mar 2024

%\cite{Ma:2010yw}
\bibitem{Ma:2010yw}
Y.~Q.~Ma, K.~Wang and K.~T.~Chao,
%``$J/\psi (\psi^\prime)$ production at the Tevatron and LHC at ${\cal O}(\alpha_s^4v^4)$ in nonrelativistic QCD,''
Phys. Rev. Lett. \textbf{106}, 042002 (2011)
%doi:10.1103/PhysRevLett.106.042002
[arXiv:1009.3655 [hep-ph]].
%288 citations counted in INSPIRE as of 09 Mar 2024

%\cite{Zhang:2009ym}
\bibitem{Zhang:2009ym}
Y.~J.~Zhang, Y.~Q.~Ma, K.~Wang and K.~T.~Chao,
%``QCD radiative correction to color-octet $J/\psi$ inclusive production at B Factories,''
Phys. Rev. D \textbf{81}, 034015 (2010)
%doi:10.1103/PhysRevD.81.034015
[arXiv:0911.2166 [hep-ph]].
%102 citations counted in INSPIRE as of 09 Mar 2024

%\cite{Gong:2012ug}
\bibitem{Gong:2012ug}
B.~Gong, L.~P.~Wan, J.~X.~Wang and H.~F.~Zhang,
%``Polarization for Prompt J/\ensuremath{\psi} and \ensuremath{\psi}(2s) Production at the Tevatron and LHC,''
Phys. Rev. Lett. \textbf{110}, no.4, 042002 (2013)
%doi:10.1103/PhysRevLett.110.042002
[arXiv:1205.6682 [hep-ph]].
%233 citations counted in INSPIRE as of 09 Mar 2024

%\cite{Brambilla:2024iqg}
\bibitem{Brambilla:2024iqg}
N.~Brambilla, M.~Butenschoen and X.~P.~Wang,
%``How well does nonrelativistic QCD factorization work at next-to-leading order?,''
[arXiv:2411.16384 [hep-ph]].
%2 citations counted in INSPIRE as of 02 Jun 2025

%\cite{CMS:2013gbz}
\bibitem{CMS:2013gbz}
S.~Chatrchyan \textit{et al.} [CMS],
%``Measurement of the Prompt $J/\psi$ and $\psi$(2S) Polarizations in $pp$ Collisions at $\sqrt{s}$ = 7 TeV,''
Phys. Lett. B \textbf{727}, 381-402 (2013)
%doi:10.1016/j.physletb.2013.10.055
[arXiv:1307.6070 [hep-ex]].
%213 citations counted in INSPIRE as of 20 Feb 2025

%\cite{CMS:2024igk}
\bibitem{CMS:2024igk}
A.~Hayrapetyan \textit{et al.} [CMS],
%``Measurement of the polarizations of prompt and non-prompt Image 1 and \ensuremath{\psi}(2S) mesons produced in pp collisions at s=13TeV,''
Phys. Lett. B \textbf{858}, 139044 (2024)
%doi:10.1016/j.physletb.2024.139044
[arXiv:2406.14409 [hep-ex]].
%2 citations counted in INSPIRE as of 20 Feb 2025

%\cite{Qiao:2003ba}
\bibitem{Qiao:2003ba}
C.~F.~Qiao and J.~X.~Wang,
%``$J/\psi$ + $c$ + $\bar{c}$ photoproduction in $e^{+} e^{-}$ scattering,''
Phys. Rev. D \textbf{69}, 014015 (2004)
%doi:10.1103/PhysRevD.69.014015
[arXiv:hep-ph/0308244 [hep-ph]].
%27 citations counted in INSPIRE as of 21 May 2025

%\cite{Zhang:2006ay}
\bibitem{Zhang:2006ay}
Y.~J.~Zhang and K.~T.~Chao,
%``Double charm production e+ e- ---\ensuremath{>} J / psi + c anti-c at B factories with next-to-leading order QCD correction,''
Phys. Rev. Lett. \textbf{98}, 092003 (2007)
%doi:10.1103/PhysRevLett.98.092003
[arXiv:hep-ph/0611086 [hep-ph]].
%113 citations counted in INSPIRE as of 21 May 2025

%\cite{Chen:2016hju}
\bibitem{Chen:2016hju}
Z.~Q.~Chen, L.~B.~Chen and C.~F.~Qiao,
%``NLO QCD Corrections for $J/\psi+ c + \bar{c}$ Production in Photon-Photon Collision,''
Phys. Rev. D \textbf{95}, no.3, 036001 (2017)
%doi:10.1103/PhysRevD.95.036001
[arXiv:1608.06231 [hep-ph]].
%13 citations counted in INSPIRE as of 21 May 2025

%\cite{Feng:2024heh}
\bibitem{Feng:2024heh}
Q.~M.~Feng and C.~F.~Qiao,
%``NLO corrections to J/\ensuremath{\psi}+c+c\textasciimacron{} photoproduction,''
Phys. Rev. D \textbf{110}, no.9, 094047 (2024)
%doi:10.1103/PhysRevD.110.094047
[arXiv:2405.05683 [hep-ph]].
%1 citations counted in INSPIRE as of 21 Feb 2025

%\cite{Braaten:2014ata}
\bibitem{Braaten:2014ata}
E.~Braaten and J.~Russ,
%``$J/\psi$ and $\Upsilon$ Polarization in Hadronic Production Processes,''
Ann. Rev. Nucl. Part. Sci. \textbf{64}, 221-246 (2014)
%doi:10.1146/annurev-nucl-030314-044352
[arXiv:1401.7352 [hep-ex]].
%12 citations counted in INSPIRE as of 18 May 2025

%\cite{Faccioli:2010kd}
\bibitem{Faccioli:2010kd}
P.~Faccioli, C.~Lourenco, J.~Seixas and H.~K.~Wohri,
%``Towards the experimental clarification of quarkonium polarization,''
Eur. Phys. J. C \textbf{69}, 657-673 (2010)
%doi:10.1140/epjc/s10052-010-1420-5
[arXiv:1006.2738 [hep-ph]].
%215 citations counted in INSPIRE as of 18 May 2025

%\cite{Mastrolia:2012bu}
\bibitem{Mastrolia:2012bu}
P.~Mastrolia, E.~Mirabella and T.~Peraro,
%``Integrand reduction of one-loop scattering amplitudes through Laurent series expansion,''
JHEP \textbf{06}, 095 (2012)
[erratum: JHEP \textbf{11}, 128 (2012)]
%doi:10.1007/JHEP11(2012)128
[arXiv:1203.0291 [hep-ph]].
%153 citations counted in INSPIRE as of 25 Apr 2024

%\cite{Peraro:2014cba}
\bibitem{Peraro:2014cba}
T.~Peraro,
%``Ninja: Automated Integrand Reduction via Laurent Expansion for One-Loop Amplitudes,''
Comput. Phys. Commun. \textbf{185}, 2771-2797 (2014)
%doi:10.1016/j.cpc.2014.06.017
[arXiv:1403.1229 [hep-ph]].
%141 citations counted in INSPIRE as of 09 Mar 2024

%\cite{Wu:2023upw}
\bibitem{Wu:2023upw}
Z.~Wu, J.~Boehm, R.~Ma, H.~Xu and Y.~Zhang,
%``NeatIBP 1.0, a package generating small-size integration-by-parts relations for Feynman integrals,''
Comput. Phys. Commun. \textbf{295}, 108999 (2024)
%doi:10.1016/j.cpc.2023.108999
[arXiv:2305.08783 [hep-ph]].
%14 citations counted in INSPIRE as of 09 Mar 2024

%\cite{Harris:2001sx}
\bibitem{Harris:2001sx}
B.~W.~Harris and J.~F.~Owens,
%``The Two cutoff phase space slicing method,''
Phys. Rev. D \textbf{65}, 094032 (2002)
%doi:10.1103/PhysRevD.65.094032
[arXiv:hep-ph/0102128 [hep-ph]].
%405 citations counted in INSPIRE as of 21 Feb 2025

%\cite{Dittmaier:2003bc}
\bibitem{Dittmaier:2003bc}
S.~Dittmaier,
%``Separation of soft and collinear singularities from one loop N point integrals,''
Nucl. Phys. B \textbf{675}, 447-466 (2003)
%doi:10.1016/j.nuclphysb.2003.10.003
[arXiv:hep-ph/0308246 [hep-ph]].
%147 citations counted in INSPIRE as of 10 Mar 2024

%\cite{Pumplin:2002vw}
\bibitem{Pumplin:2002vw}
J.~Pumplin, D.~R.~Stump, J.~Huston, H.~L.~Lai, P.~M.~Nadolsky and W.~K.~Tung,
%``New generation of parton distributions with uncertainties from global QCD analysis,''
JHEP \textbf{07}, 012 (2002)
%doi:10.1088/1126-6708/2002/07/012
[arXiv:hep-ph/0201195 [hep-ph]].
%6982 citations counted in INSPIRE as of 14 Mar 2024

%\cite{ParticleDataGroup:2020ssz}
\bibitem{ParticleDataGroup:2020ssz}
P.~A.~Zyla \textit{et al.} [Particle Data Group],
%``Review of Particle Physics,''
PTEP \textbf{2020}, no.8, 083C01 (2020)
%doi:10.1093/ptep/ptaa104
%3266 citations counted in INSPIRE as of 09 Apr 2022

%\cite{ATLAS:2023qnh}
\bibitem{ATLAS:2023qnh}
G.~Aad \textit{et al.} [ATLAS],
%``Measurement of the production cross-section of $J/\psi $ and $\psi (2{\textrm{S}})$ mesons in pp collisions at $\sqrt{s} = 13$~TeV with the ATLAS detector,''
Eur. Phys. J. C \textbf{84}, no.2, 169 (2024)
%doi:10.1140/epjc/s10052-024-12439-9
[arXiv:2309.17177 [hep-ex]].
%6 citations counted in INSPIRE as of 18 May 2025

%\cite{ATLAS:2014ala}
\bibitem{ATLAS:2014ala}
G.~Aad \textit{et al.} [ATLAS],
%``Measurement of $\chi_{c1}$ and $\chi_{c2}$ production with $\sqrt{s}$ = 7 TeV $pp$ collisions at ATLAS,''
JHEP \textbf{07}, 154 (2014)
%doi:10.1007/JHEP07(2014)154
[arXiv:1404.7035 [hep-ex]].
%100 citations counted in INSPIRE as of 18 May 2025

%\cite{Zhang:2008gp}
\bibitem{Zhang:2008gp}
Y.~J.~Zhang, Y.~Q.~Ma and K.~T.~Chao,
%``Factorization and NLO QCD correction in $e^+e^- \to J/\psi(\psi(2S))+\chi_{c0}$ at B Factories,''
Phys. Rev. D \textbf{78}, 054006 (2008)
%doi:10.1103/PhysRevD.78.054006
[arXiv:0802.3655 [hep-ph]].
%66 citations counted in INSPIRE as of 18 May 2025

%\cite{Braaten:1993mp}
\bibitem{Braaten:1993mp}
E.~Braaten, K.~m.~Cheung and T.~C.~Yuan,
%``Z0 decay into charmonium via charm quark fragmentation,''
Phys. Rev. D \textbf{48}, 4230-4235 (1993)
%doi:10.1103/PhysRevD.48.4230
[arXiv:hep-ph/9302307 [hep-ph]].
%308 citations counted in INSPIRE as of 30 Oct 2024

%\cite{Zheng:2019dfk}
\bibitem{Zheng:2019dfk}
X.~C.~Zheng, C.~H.~Chang and X.~G.~Wu,
%``NLO fragmentation functions of heavy quarks into heavy quarkonia,''
Phys. Rev. D \textbf{100}, no.1, 014005 (2019)
%doi:10.1103/PhysRevD.100.014005
[arXiv:1905.09171 [hep-ph]].
%28 citations counted in INSPIRE as of 30 Oct 2024

%\cite{Gong:2011et}
\bibitem{Gong:2011et}
B.~Gong, R.~Li and J.~X.~Wang,
%``Testing the QCD fragmentation mechanism on heavy quarkonium production at LHC,''
[arXiv:1102.0118 [hep-ph]].
%1 citations counted in INSPIRE as of 04 Jun 2025
\end{thebibliography}
\end{document}